\documentclass{elsart}
\usepackage{epsfig}

\begin{document}
\begin{frontmatter}

\title{Brownian Dynamics, Time-averaging and Colored Noise}
\author[Rio1]{D. O. Soares-Pinto\thanksref{CNPq}}
\author[Rio]{and W. A. M.  Morgado\thanksref{FAPERJCNPq}}
\address[Rio1]{Centro Brasileiro de Pesquisas
F\'{\i}sicas, \\ Rua Dr. Xavier Sigaud 150, CEP 22290-180, Rio de
Janeiro, Brazil} \address[Rio]{welles@fis.puc-rio.br \\ Departamento
de F\'{\i}sica, Pontif\'{\i}cia Universidade Cat\'olica do Rio de
Janeiro, C.P.  38071, 22452-970 Rio de Janeiro, Brazil}
\thanks[CNPq]{Partially supported by CNPq, Brazil.}
\thanks[FAPERJCNPq]{Partially supported by FAPERJ and CNPq, Brazil.}

\maketitle

\begin{abstract}
We propose a method to obtain the equilibrium distribution for
positions and velocities of a one-dimensional particle via
time-averaging and Laplace transformations. We apply it to the case of
a damped harmonic oscillator in contact with a thermal bath. The
present method allows us to treat, among other cases, a Gaussian noise
function exponentially correlated in time, e.g., Gaussian colored
noise. We obtain the exact equilibrium solution and study some of its
properties.
\end{abstract}
\end{frontmatter}


\section{Introduction}

A Brownian particle is a massive particle ($M$), surrounded by a bath
of lighter particles ($m$)~\cite{nelson}. Its motion is very irregular
and varies rapidly with time. However, its averaged behavior can be
well understood. The reason for this is that the mass ratio ($m/M\ll
1$) leads to the separation of time-scales for the slow Brownian's
degrees of freedom from the fast bath's degrees of
freedom~\cite{iovk}. This allows us to construct adequate methods
(e.g. Fokker-Planck equations) to describe the time evolution for the
probability distribution function, or equivalently, for the momenta of
the distribution~\cite{risken,vk}. The equilibrium distribution is
thus obtained as a function of the equilibrium momenta.  We do not
need to point out that the behavior of Brownian particles has been the
subject of many scientific studies since Einstein's seminal
papers~\cite{einstein,hanggi}, already a hundred years
old. Interesting extensions of this old problem have attracted the
interest of many groups lately~\cite{brown}. The treatment of
non-standard forms of the noise function has also been successfully
obtained recently~\cite{dubkov}.

In this paper, we calculate the equilibrium distribution for a damped
harmonically bound Brownian particle as a direct integral, by means of
Laplace transforms. However, a word is necessary to explain the use of
the heavy machinery of Laplace transforms upon such a well known
model.

In fact, the present model is only used as an illustration for the
time-averaging method we propose. Applications for more interesting
systems will follow.  It basically consists in arranging the full
dynamical behavior, expressed by the exact time-averaged distribution,
into sums of integrals that can be proved either to vanish identically
or to be easily calculable. Thus, the exact equilibrium distribution
is obtained. Further extensions of the method are under way for
non-Gaussian noise~\cite{premoc}.

Technically, we define the equilibrium distribution in the spirit of
Boltzmann's prescription~\cite{kac} as a time-averaged distribution.
The equilibrium distribution calculation is done by carrying out
analytically the calculation for the time average of the distribution
defined below, via the use of the stochastic functions $x(t)$ and
$v(t)$, solutions of the stochastic dynamics.  It reads
\begin{equation}
\mbox{P}_{eq}(x,v)=\lim_{T\rightarrow\infty}\frac{1}{T}\int_0^{T}dt\,
<\delta(x-x(t))\delta(v-v(t))>,\label{timeaverage}
\end{equation}
where $<>$ means average over the noise $\eta(t)$.
%
%
For ergodic systems fulfilling 
\[
f(t\rightarrow\infty)=\lim_{s\rightarrow0}s\tilde{f}(s),
\]
where $\tilde{f}$ is the Laplace transform of $f$, we  can 
%
%
use an equivalent form for Eq.~\ref{timeaverage}, namely
\begin{equation}
\mbox{P}_{eq}(x,v)=\lim_{z\rightarrow 0}z\int_0^{\infty}dt\, e^{-zt}
<\delta(x-x(t))\delta(v-v(t))>.\label{timeaveragez}
\end{equation}

As mentioned above, the functions $x(t)$ and $v(t)$ are solutions of
the Langevin equations that follow from coupling the Brownian particle
to a bath of lighter particles. We assume these to influence the
Brownian dynamics via an effective non-white Gaussian noise force
$\eta(t)$. The effect of time-correlations leads to the
renormalization of the oscillator frequency at equilibrium (all
results are exact), discussed in the next section. 


Our approach is straightforward and allows us to separate the
reversible from the irreversible parts of the dynamics. The
equilibrium distribution (the irreversible part) is obtained by
opening the delta-function on Eq.~\ref{timeaveragez} by Laplace
transformation into an infinite series, and by keeping only
non-vanishing terms in it (in the limit $z\rightarrow 0$).
As will be made clear in the following, the vanishing terms (in the
limit $t\rightarrow\infty\Leftrightarrow z\rightarrow0$) do so due to
their singular structure. They have poles (the same as those for the
dispersion relation of the damped harmonic oscillator described by the
Langevin Equation minus the random force) expressing the dissipation
due to the damping.  The surviving terms are the ones that do not
carry the effect of the damping to the main propagator (as shall be
defined below, proportional to $z$). The equilibrium distribution is
obtained by summing over the non-vanishing terms.

The \underline{method} we propose could be summarized as follows:
\begin{itemize}
\item Laplace transform of the Langevin equation and obtain
$\tilde{x}(z)$ and $\tilde{v}(z)$;
\item Laplace transform of the Equilibrium distribution defined by
Equation~\ref{timeaverage};
\item To eliminate the vanishing terms (in the limit $z\rightarrow0
\Leftrightarrow t\rightarrow\infty$);
\item To integrate over the the non-vanishing terms and obtain the
equilibrium distribution.
\end{itemize}
This method for obtaining the equilibrium distribution via
time-averages, can be extended to different systems~\footnote{Work is
under way in order to extend the present results to more general forms
of noise, in special for the case of non-Gaussian white noise.}.

As stressed above, an important aspect of the method is that it allows
us to study time-dependent processes (e.g., equilibration), since the
effect of time comes from a single propagator. The long-time behavior
of any dynamical functions for the system are easily singled out in
the limit $z\rightarrow 0 \Leftrightarrow t\rightarrow\infty$.

This paper is organized as follows. In Section 2 we define the
model. In Sections 3 we define, and Laplace transform, the
time-averaged equilibrium distribution. In Section 4 we discuss the
irreversibility from the point of view of surviving and vanishing
integrals. In Section 5 we select the non-vanishing contributions and
obtain the equilibrium distribution in an analytical form. In section
6 we make some concluding remarks.

\section{Model}

\subsection{Langevin Equations and Laplace transforms}
We model a Brownian particle subject to a damped harmonic
potential. The damping is proportional to the velocity. A Langevin
force $\eta(t)$, Gaussian distributed, effectively acts on the
Brownian particle due to the interaction with the bath particles,
not included in the dynamics. We assume that $\eta(t)$ can be
successfully Laplace transformed into $\tilde{\eta}(z)$.

The system obeys the following coupled Langevin-like
Equations~\footnote{Eq.~\ref{langevin} is strictly of Langevin type
when the noise $\eta(t)$ is white. A discussion on this can be found
in reference~\cite{vk}, chapter IX.}:
\begin{equation}\label{langevin}
m\dot{v}(t)=-{\gamma}v(t)-kx(t)+{\eta}(t),
\end{equation}
\begin{equation}\label{langevin2}
\dot{x}(t)=v(t),
\end{equation}
where $\eta(t)$ is a colored noise (time-correlated) distributed
stochastic force.

The initial conditions are given by
\begin{displaymath}
x(t=0)=x_{0},\,\,
v(t=0)=v_{0}.
\end{displaymath}

Integrating Eqs.~\ref{langevin} and~\ref{langevin2}, we obtain that
$x(t)$ and $v(t)$ can be expressed respectively as:
\begin{equation}\label{xdot}
x(t)=x_{0}+\int_{0}^{t}v(t')dt',
\end{equation}
and
\begin{equation}\label{vdot}
mv(t)=mv_{0}-{\gamma}\int_{0}^{t}v(t')dt'-kx_{0}t-k
\int_{0}^{t}\int_{0}^{t'}v(t'')dt''dt'+
\int_{0}^{t}{\eta}(t')dt'.
\end{equation}

The Laplace transform of $f(t)$ is defined as
\begin{equation}
\tilde{f}(z)=\int_{0}^{\infty}dt \, e^{-zt}f(t),
\end{equation}
with Re$(z)>0.$

By taking the Laplace transform of Equations~\ref{xdot}
and~\ref{vdot}, we obtain:
\begin{equation}
\tilde{x}(z)=A(z)+B(z)\tilde{\eta}(z),
\end{equation}
and
\begin{equation}
\tilde{v}(z)=C(z)+D(z)\tilde{\eta}(z),
\end{equation}
where
\begin{equation}
A(z)=\frac{(z+{\beta})x_{0}+v_{0}}{z(z+{\beta})+w_{0}^{2}},
\end{equation}
\begin{equation}
B(z)=\frac{m^{-1}}{z(z+{\beta})+w_{0}^{2}},
\end{equation}
\begin{equation}
C(z)=\frac{zv_{0}-w_{0}^{2}x_{0}}{z(z+{\beta})+w_{0}^{2}},
\end{equation}
\begin{equation}
D(z)=\frac{m^{-1}z}{z(z+{\beta})+w_{0}^{2}},
\end{equation}
where $w_{0}^{2}=\frac{k}{m}$ and ${\beta}=\frac{\gamma}{m}$.

Let's observe that only the coefficients $A$ and $C$ carry the
contributions from the initial conditions. They are the memory terms:
the vanishing of their contribution will make the process
irreversible.

The common denominator of the coefficients $A$, $B$, $C$ and $D$ is
the damped harmonic oscillator's dispersion relation. Its poles are
\begin{eqnarray}
\kappa_{\pm}& =& -\frac{\beta}{2}\pm
i\frac12\sqrt{4\omega_0^2-\beta^2},\label{poles}
\end{eqnarray}
where we assume that $4\omega^2_0-\beta^2>0$. This will be
justified in the sequence, by the end of Section 5.

We observe that the coefficients $A$ and $C$ carry information about
the initial conditions, but not $B$ and $D$, which are related to the
stochastic part of the dynamics. The system will retain memory of its
initial condition as far as terms depending on the coefficients $A$
and $C$ contribute to the final results.

\subsection{Colored noise $\eta(t)$}

We assume that the stochastic process $\eta(t)$ is of the form of
Gaussian colored noise~\cite{vk}, in which $\eta(t)$ is uncorrelated
for $t<0$. It is more general than white noise and reduces to it in
the limit $\tau\rightarrow0$.

%
%
Some considerations have to be made about our choice of colored
noise. For a closed system, noise has only internal causes. Using
projection operator techniques~\cite{po} one obtains a generalized
Langevin Equation~\cite{mazur} showing that the
fluctuation-dissipation theorem holds near equilibrium and has to be
consistent with a retarded dissipation of the form:
\[
\dot{v}(t)=-\int_{0}^{t}dt'\, \phi(t-t')v(t') + \eta(t),
\]
where
\[
\phi(t)=\frac{\left<\eta(t)\eta(t+\tau)\right>}{k_BT}.
\]
However, we are assuming that the much simpler form of dissipation
holds, namely Eq.~\ref{langevin}. This can be justified by assuming
that in a general situation (far from equilibrium), the
fluctuation-dissipation theorem does not apply and the dissipation's
memory kernel and the fluctuation properties of the noise need not to
be related~\cite{budini}. A method for solving the fully retarded
dissipation for arbitrary noise can be found in
reference~\cite{budini}.

Another interesting situation is the case of purely external noise. In
this case, noise is not described by a closed Hamiltonian and
dissipation and noise fluctuations are not related.~\footnote{We could
imagine a simple system consisting of a charged particle immersed in a
viscous fluid, driven by a strong external fluctuating electrical
field.}
%
%

Thus, we start from Eqs.~\ref{langevin} and~\ref{langevin2} and study
the long-time behavior of a non-Markovian coupled system ($x,v$). As
for the noise properties, due to its Gaussian nature, only the second
order cumulant survives (zero average distributed):
\begin{equation}
\langle\eta(t)\rangle=0;
\end{equation}
\begin{equation}
\langle\eta(t)\eta(t^{'})\rangle=\frac{D}{2\tau}
e^{-|t-t^{'}|/\tau}.
\end{equation}
As usual, for even $l$ we have:
\begin{eqnarray*}
\langle\prod_{i=1}^{l}\eta(t_i)\rangle &=&
 \sum^{\mbox{all pairwise}}_{\mbox{arrangements}}
\langle\eta(t_{i_1})\eta(t_{i_2})\rangle\ldots\langle\eta(t_{i_{l-1}})
\eta(t_{i_l})\rangle,
\end{eqnarray*}
for all possible combinations for $l=2p$ even (it will have
$\frac{(2p)!}{2^{p}p!}$ combinations of pairs of ${\eta}$
functions).

By Laplace transforming, the $\eta$-average second cumulant reads:
\begin{eqnarray}
\langle\tilde{\eta}(iq_i+\epsilon)\tilde{\eta}(iq_j+\epsilon)\rangle &=&
\left\{\frac{D}{\left[i(q_i+q_j)+2\epsilon\right]
\left[1-\tau(iq_i+\epsilon)\right]
\left[1-\tau(iq_j+\epsilon)\right]}\right\}-\nonumber\\
&&\hspace{-4.0cm}-\frac{D\tau}{2}\left\{\frac{3+
\tau\left[i(q_i+q_j)+2\epsilon\right]
-\tau^2(iq_i+\epsilon)(iq_j+\epsilon)}{
\left[1-\tau(iq_i+\epsilon)\right]\left[1+\tau(iq_i+\epsilon)\right]
\left[1-\tau(iq_j+\epsilon)\right]
\left[1+\tau(iq_j+\epsilon)\right]}\right\}.\label{etaprop}
\end{eqnarray}

Eq.~\ref{etaprop} reduces to the white noise case
\[
\frac{D}{i(q_i+q_j)+2\epsilon},
\]
as $\tau\rightarrow0$.

The presence of time-correlated noise will affect the equilibrium
distribution. As shall be seen, it has the effect of renormalizing
the oscillator frequency. The white-noise equilibrium form is
recovered when $\tau\rightarrow0$.

\section{Equilibrium Distribution}
A convenient way of defining the long-time limit
($t\rightarrow\infty$) for a convergent equilibrium distribution is to use
\begin{equation}
f^z_{eq}=\lim_{z\rightarrow0} \frac{\int_{0}^{\infty}dt\,
e^{-zt}f(t)}
{\int_{0}^{\infty}e^{-zt}dt}
\end{equation}
for $z > 0$.
This is equivalent to
\[
f_{eq}=\lim_{T\rightarrow\infty}\frac{1}{T}\int_{0}^{T}dt \,f(t),
\]
whenever this expression converges. In this case we write
\[
f(t\rightarrow \infty)=f_{eq}+\Delta(t),
\]
with $\lim_{t\rightarrow\infty}\Delta(t)=0$\footnote{We assume
that: $\forall \delta>0, \exists M:\forall t> M\Rightarrow
\left|\Delta(t)\right|<\delta.$}. For $z=\frac{1}{T}$, the average
$f^z_{eq}$ then reads
\begin{eqnarray*}
f^z_{eq}&=&\lim_{z\rightarrow0} \frac{\int_{0}^{\infty}dt\,
e^{-zt}f(t)}
{\int_{0}^{\infty}e^{-zt}dt}
=\lim_{T\rightarrow\infty} \frac{\int_{0}^{\infty}dt\,
e^{-t/T}f(t)}{T}\\
&=&f_{eq}+
\lim_{T\rightarrow\infty} \frac{\int_{0}^{M}dt\,
e^{-t/T}\Delta(t)}{T}
\\
\Rightarrow \lim_{\delta\rightarrow0}\left|f^z_{eq}-f_{eq}\right|&\leq&
\lim_{\delta\rightarrow0}\delta\rightarrow 0.
\end{eqnarray*}
The two forms of time averaging are equivalent. In the case the
distribution defined on Equation~\ref{timeaverage} does not converge
to an analytical form (as $t\rightarrow\infty$), both definitions
above will fail to give convergent values.


Thus, we write the time-averaged equilibrium distribution as (see
Appendix A):
\begin{eqnarray}
P_{eq}(x,v)&=&\lim_{z\rightarrow0}z
\int_{0}^{\infty}dt \,
e^{-zt}\langle{\delta}(x-x(t)){\delta}(v-v(t))\rangle
\nonumber\\
&=&\lim_{z\rightarrow0}\lim_{\epsilon\rightarrow0}
\sum_{l,m=0}^{\infty}
\int_{-\infty}^{+\infty}\frac{dQ}{2\pi}\frac{dP}{2\pi}e^{iQx+iPv}
\frac{(-iQ)^{l}}{l!}\frac{(-iP)^{m}}{m!}
\nonumber\\
&\times&\int_{-\infty}^{+\infty}
\prod_{f=1}^{l}\frac{dq_{f}}{2\pi}\prod_{h=1}^{m}\frac{dp_{h}}{2\pi}
\frac{z}{z-\left[\sum_{f=1}^{l}iq_{f}+
\sum_{h=1}^{m}ip_{h}+(l+m)\epsilon\right]}
\nonumber\\
&\times&
\langle\prod_{f=1}^{l}\tilde{x}(iq_{f}+\epsilon)\prod_{h=1}^{m}
\tilde{v}(ip_{h}+\epsilon)\rangle.\label{Peqexp}
\end{eqnarray}
The integration path for the $q$- and $p$-variables is also given in
Figure~\ref{fig1}. Both upper and lower semi-hemispherical integration
paths will contribute nothing to the total integral. We choose the upper
path for our calculations.

\section{Origin of irreversibility at the limit ${\it z}\rightarrow
0$}

The equilibrium distribution we obtain corresponds to a sum of
(infinite) products of polynomial fractions with poles given by the
dispersion relation, Equation~\ref{poles}. One of those propagators,
in Equation~\ref{Peqexp}, stands out for being the only one that
contains the effect of time: the main propagator $G(z)$, given by
\begin{equation}
G(z)= \frac{z}
{z-[i(q_{1}+\ldots+q_{l}+p_{1}+\ldots+p_{m})+(l+m)\epsilon]}.\label{MP}
\end{equation}
In order to evaluate the equilibrium distribution (the analysis can
also be extended to the non-equilibrium case $z\neq0$), we need to
study the poles and the limiting properties of the main propagator
given by Equation~\ref{MP}.

When integrating a given $q$ or $p$--variable, by choosing the upper path
on the complex plane shown in figure~\ref{fig1}, the pole associated
with the propagator falls outside the integration path, contributing
with no residues. However, after effecting all $(m+l)$ variable
calculations, values associated with the poles of the the rest of the
integrand in Eq.~\ref{Peqexp} might be carried through into the
denominator of $G(z)$.  It is essential for us to analyze $G(z)$ at
the limit when $z\rightarrow0$.

When the denominator in the main propagator is non-zero for
$z\rightarrow 0$, it is clear that $\lim_{z\rightarrow0}G(z) = 0$. A
non-zero result can only be obtained, in the limit $z\rightarrow 0$,
when the sum of $q$ and $p$--variable is completely eliminated from
$G(z)$. When all variables are eliminated, we have instead
$\lim_{z\rightarrow0}G(z) = \lim_{z\rightarrow0}\frac{z}{z}=1$, and
the integral yields a non-zero contribution.

Since the factors $A$ and $C$ are not coupled with the stochastic
force $\tilde{\eta}$, the $q$ and $p$--variables associated with
them will lead to residues only around the poles of the dispersion
relation ($-i\kappa_{\pm}$). The pole's values will in the end be
transported all the way to the denominator of the main propagator.
So, by integrating over the poles of the functions $A$ and $C$, a
non-zero final value to the summation $i\left[q_{1}+\ldots+q_{l}
+p_{1} +\ldots+ p_{m}\right]+(m+l)\epsilon$, will give a zero
contribution in the limit $z\rightarrow 0$.

This is the simple way by which irreversibility manifests itself on
the path to equilibrium: the memory of the initial conditions is lost
as the inverse Laplace transform of terms such as
\[
\sim \frac{z}{z-i(m\,\,(-i\kappa_{\pm}))}=\frac{z}{z-m\beta/2\pm
i\ldots}
\rightarrow \,\,\,
 e^{-\frac{m\beta t}{2}}\rightarrow 0,
\]
where $m$ is an integer.

\section{Important integrals}
The effect of poles from the dispersion relation in functions $B$ and
$D$ can be eliminated since they are coupled with
$\tilde{\eta}$-functions, which are associated with a factor
$\delta(iq+iq^{'} + 2\epsilon)$ dependency, coming from the stochastic
properties of the $\tilde{\eta}$-functions. When we integrate over the
$q$ and $q^{'}$ variables, the $\delta(iq+iq^{'} + 2\epsilon)$
dependency eliminates the effect of these variables from the main
propagator in the equilibrium limit $z\rightarrow0$. Thus, only three
possible couplings may contribute to the equilibrium distribution:
$BB$, $DD$, and $BD$.

In the following we list the non-vanishing contributions to the
equilibrium distribution P$_{eq}(x,v)$.

\subsection{Contribution from $BB$}
The typical contributing term for the $x$ distribution is given by the
$BB$ contributing terms below (for details see Appendix B):

\begin{displaymath}
\int_{-\infty}^{+\infty}\frac{dq_{i}}{2\pi}\frac{dq_{j}}{2\pi}
\frac{z}{z-i(q_{i}+q_{j}+\diamond)}B(iq_{i}+\epsilon)B(iq_{j}+\epsilon)
\langle\tilde{\eta}(iq_{i}+\epsilon)\tilde{\eta}(iq_{j}+\epsilon)\rangle=
\end{displaymath}
\begin{equation}
=\frac{z}{z-i\diamond}\frac{D}{2[m+{\tau}^{2}k-{\tau}{\gamma}][m+{\tau}^{2}k+{\tau}{\gamma}]}\{\frac{m^{2}}{k\gamma}-\frac{{\tau}^{2}{\gamma}}{k}+\frac{{\tau}^{2}m}{\gamma}+{\tau}^{3}\}
\end{equation}
where $\diamond$ represents all the non integrated variables.

Consequently, the only contributing terms will be even powers $m =
2\theta$:
\begin{displaymath}
\frac{m!}{2^{\frac{m}{2}}(\frac{m}{2})!}\{\frac{1}{[(m+{\tau}^{2}k)^{2}-{\tau}^{2}{\gamma}^{2}]}[\frac{Dm^{2}}{2k\gamma}-\frac{{\tau}^{2}{D\gamma}}{2k}+\frac{{\tau}^{2}Dm}{2\gamma}+\frac{{\tau}^{3}D}{2}]\}^{\frac{m}{2}}=
\end{displaymath}
\begin{displaymath}
=\frac{(2\theta)!}{2^{\theta}{\theta}!}\{\frac{1}{[(m+{\tau}^{2}k)^{2}-{\tau}^{2}{\gamma}^{2}]}[\frac{Dm^{2}}{2k\gamma}-\frac{{\tau}^{2}{D\gamma}}{2k}+\frac{{\tau}^{2}Dm}{2\gamma}+\frac{{\tau}^{3}D}{2}]\}^{\theta}
\end{displaymath}

\subsection{Contribution from $DD$}
Similarly with the $BB$, the $DD$ contributing terms are:

\begin{displaymath}
\int_{-\infty}^{+\infty}\frac{dp_{i}}{2\pi}\frac{dp_{j}}{2\pi}
\frac{z}{z-i(p_{i}+p_{j}+\diamond)}D(ip_{i}+\epsilon)D(ip_{j}+\epsilon)
\langle\tilde{\eta}(ip_{i}+\epsilon)\tilde{\eta}(ip_{j}+\epsilon)\rangle=
\end{displaymath}
\begin{equation}
=\frac{z}{z-i\diamond}\frac{-D}{2[m+{\tau}^{2}k-{\tau}{\gamma}][m+{\tau}^{2}k+{\tau}{\gamma}]}[\frac{-m}{\gamma}-\frac{{\tau}^{2}k}{\gamma}+{\tau}]
\end{equation}

In the limit $z\rightarrow0$, the contribution from the velocity momenta
averages for the probability distribution is given by, for $n=2\alpha$:

\begin{displaymath}
\frac{n!}{2^{\frac{n}{2}}(\frac{n}{2})!}\{\frac{1}{[(m+{\tau}^{2}k)^{2}-{\tau}^{2}{\gamma}^{2}]}[\frac{Dm}{2\gamma}+\frac{{\tau}^{2}Dk}{2\gamma}-\frac{{\tau}D}{2}]\}^{\frac{n}{2}}=
\end{displaymath}
\begin{displaymath}
=\frac{(2\alpha)!}{2^{\alpha}{\alpha}!}\{\frac{1}{[(m+{\tau}^{2}k)^{2}-{\tau}^{2}{\gamma}^{2}]}[\frac{Dm}{2\gamma}+\frac{{\tau}^{2}Dk}{2\gamma}-\frac{{\tau}D}{2}]\}^{\alpha}
\end{displaymath}

\subsection{Contribution from $BD$}
Similarly with the above, we can calculate the cross contribution $BD$
except that it has an odd power of the integrated variables (${\it
O}(p)$) in the numerator while the previous ones had even powers of it
(${\it O}(1)$ and ${\it O}(p^2)$ respectively). Thus:
\begin{equation}
\int_{-\infty}^{+\infty}\frac{dq_{i}}{2\pi}\frac{dp_{j}}{2\pi}\frac{z}{z-i(q_{i}+p_{j}+\diamond)}
B(iq_{i}+\epsilon)D(ip_{j}+\epsilon)\langle\tilde{\eta}(iq_{i}+\epsilon)\tilde{\eta}(ip_{j}+\epsilon)\rangle,
\end{equation}
the only important term
\begin{equation}
\int_{-\infty}^{+\infty}\frac{dp_{j}}{2\pi}\frac{i
z}{z-i(0+\diamond)}
\frac{m^{-2}Dp_{j}}{(p_{j}-i\kappa_{+})(p_{j}-i\kappa_{-})(p_{j}+i\kappa_{+})(p_{j}+i\kappa_{-})}
\end{equation}
is identically null because of the odd power of $p_j$. So the
equilibrium distribution will be separated into a position term and a
velocity term:
\[
P_{eq}(x,v)=P_{eq}(x)P_{eq}(v).
\]

\subsection{Equilibrium distribution}

Collecting all the non-vanishing terms in the $z\rightarrow 0$ limit,
we write Eq.~\ref{Peqexp} as:

\begin{displaymath}
P_{eq}(x,v)=\sum_{\theta=0}^{\infty}\sum_{\alpha=0}^{\infty}
\int_{-\infty}^{+\infty}\frac{dQ}{2\pi}\frac{dP}{2\pi}
e^{iQx+iPv}\times
\end{displaymath}
\begin{displaymath}
\times\frac{(-iQ)^{2\theta}}{2\theta!}
\frac{(2\theta)!}{2^{\theta}{\theta}!}\{\frac{1}{[(m+{\tau}^{2}k)^{2}-{\tau}^{2}{\gamma}^{2}]}[\frac{Dm^{2}}{2k\gamma}-\frac{{\tau}^{2}{D\gamma}}{2k}+\frac{{\tau}^{2}Dm}{2\gamma}+\frac{{\tau}^{3}D}{2}]\}^{\theta}\times
\end{displaymath}
\begin{displaymath}
\times\frac{(-iP)^{2\alpha}}{2\alpha!}
\frac{(2\alpha)!}{2^{\alpha}{\alpha}!}\{\frac{1}{[(m+{\tau}^{2}k)^{2}-{\tau}^{2}{\gamma}^{2}]}[\frac{Dm}{2\gamma}+\frac{{\tau}^{2}Dk}{2\gamma}-\frac{{\tau}D}{2}]\}^{\alpha}=
\end{displaymath}
\begin{displaymath}
=\int_{-\infty}^{+\infty}\frac{dQ}{2\pi}\exp\{iQx-\frac{{Q}^{2}}{2[(m+{\tau}^{2}k)^{2}-{\tau}^{2}{\gamma}^{2}]}[\frac{Dm^{2}}{2k\gamma}-\frac{{\tau}^{2}{D\gamma}}{2k}+\frac{{\tau}^{2}Dm}{2\gamma}+\frac{{\tau}^{3}D}{2}]\}\times
\end{displaymath}
\begin{displaymath}
\times\int_{-\infty}^{+\infty}\frac{dP}{2\pi}\exp\{iPv-\frac{P^{2}}{2[(m+{\tau}^{2}k)^{2}-{\tau}^{2}{\gamma}^{2}]}[\frac{Dm}{2\gamma}+\frac{{\tau}^{2}Dk}{2\gamma}-\frac{{\tau}D}{2}]\}
\end{displaymath}

Integrating the equation above, we finally obtain the properly
normalized equilibrium distribution:

\begin{eqnarray}
P_{eq}(x,v) &=& \sqrt{
\frac{[(m+{\tau}^{2}k)^{2}-{\tau}^{2}{\gamma}^{2}]}{\pi
[\frac{Dm^{2}}{k\gamma}-\frac{{\tau}^{2}{D\gamma}}{k}+
\frac{{\tau}^{2}Dm}{\gamma}+{\tau}^{3}D]}}
\exp\{-\frac{m^2[(1+{\tau}^{2}\omega_0^2)^{2}
-{\tau}^{2}{\beta}^{2}]x^{2}}{[\frac{Dm^{2}}{k\gamma}
-\frac{{\tau}^{2}{D\gamma}}{k}+\frac{{\tau}^{2}Dm}{\gamma}
+{\tau}^{3}D]}\}\times\nonumber\\
\label{Peqfin}
&\times&\sqrt{\frac{[(m+{\tau}^{2}k)^{2}-{\tau}^{2}{\gamma}^{2}]}
{\pi [\frac{Dm}{\gamma}+\frac{{\tau}^{2}Dk}{\gamma}-{\tau}D]}}
\exp\{-\frac{m^2[(1+{\tau}^{2}\omega_0^2)^{2}-{\tau}^{2}{\beta}^{2}]
{v}^{2}}{[\frac{Dm}{\gamma}-{\tau}D+\frac{{\tau}^{2}Dk}{\gamma}]}\}
\end{eqnarray}

On the limit ${\tau}\rightarrow 0$ we obtain the distribution of
displacement and velocity of a Brownian particle under a Gaussian
white noise as expected:

\begin{equation}\label{Peqfin2}
P_{eq}(x,v)=\sqrt{\frac{k\gamma}{\pi D}}
e^{-\frac{2{\gamma}}{D}\frac{kx^2}{2}}\sqrt{\frac{m\gamma}{\pi D}}
e^{-\frac{2{\gamma}}{D}\frac{m{v}^{2}}{2}}
\end{equation}
Eq.~\ref{Peqfin2} is the analytical time-averaged equilibrium
distribution for the damped harmonically bound Brownian particle.

%
%
The result from Eq.~\ref{Peqfin} has to be compared with exact results
in the literature. The exact solution for a Langevin equation with
time-correlated noise has been obtained by means of functional
methods~\cite{caceresexato,sancho} or by path-integral
approaches~\cite{wio}. In order to compare our results with these
references, we allow the system to be unconfined (we assume
Eq.~\ref{Peqfin} holds for $k\rightarrow 0$ and $m=1$) and we
integrate over the coordinate variable $x$. The reduced distribution
obtained from Eq.~\ref{Peqfin} then reads
\begin{eqnarray}
P_{eq}(v) &=& \sqrt{\frac{\gamma[1+\tau\gamma]}
{\pi D}}\exp\{-\frac{\gamma[1+\tau\beta]v^2}{D}\}\nonumber\\
&=&  \left(\sqrt{\frac{\gamma}
{\pi D}}\exp\{-\frac{\gamma v^2}{D}\}\right)
 \left(1+\tau\left[\frac{\gamma}{2}-
\frac{\gamma^2v^2}{D}\right]+{\it O}(\tau^2)\right)
\end{eqnarray}
This corresponds exactly (up to linear order on $\tau$) to the results
in references~\cite{sancho,wio} when we make the change $D\rightarrow
2D$ due to slightly different definitions of the noise
time-correlation. As for the result in reference~\cite{caceresexato},
they are identical to all orders of $\tau$.

A difference between our approach and these other
authors'~\cite{caceresexato,sancho,wio} comes from allowing another
set of coupled variables (position $\left\{x\right\}$) to interact
with our variable (velocity $\left\{v\right\}$). This leads to the
appearance of terms proportional to $k\tau^2$ on the equilibrium
distribution. As somewhat advanced in the Appendix A of
reference~\cite{wio}, the equilibrium distribution only depends on
simple combinations of small powers of $\tau$ (on the exponent's
argument).
%
%

The fluctuation-dissipation ratio $\frac{D}{2\gamma}$ plays the role
of a thermal-bath temperature in the limit $\gamma,D\rightarrow0$ with
$D/2\gamma=$ constant:
\[
k_BT_{bath}=\frac{D}{2\gamma}.
\]
This is the reason we choose $4\omega_0^2>\beta^2$ in Section 2.1.

However, for non-zero $\tau$ the expression for the temperature grows
a little more complicated. It will be given by
\begin{eqnarray}\label{Tcolored}
k_BT_{bath}^{col-noise}&=& \frac{D[1-{\tau\beta}+{\tau}^{2}\omega_0^2]}
{2\gamma[(1+{\tau}^{2}\omega_0^2)^{2}-{\tau}^{2}{\beta}^{2}]}.
\end{eqnarray}

From the form for the temperature and the equilibrium distribution,
Eqs.~\ref{Peqfin2} and~\ref{Tcolored}, we observe that the equilibrium
frequency is renormalized to $\omega_1$, given exactly by
\begin{eqnarray}\label{Omegacolored}
\omega_1^2&=&\omega_0^2 
\frac{[1-{\tau\beta}+{\tau}^{2}\omega_0^2]}
{[1-\tau^{2}\beta^2+\tau^2\omega_0^2+\tau^3\omega_0^2\beta]}.
\end{eqnarray}

As can be seen, in first order in $\tau$ this corresponds to 
\[
\omega_1^2\approx\omega_0^2(1-{\tau\beta}),
\]
which reflects a decrease in the oscillating frequency due to
damping. The time-correlations in $\eta(t)$ tend to make the
velocities more persistent when they are highest (near the equilibrium
position) thus making the system getting further away from the origin
and taking longer to come back.

We are presently generalizing this method in order to treat other
forms of potential gradients affecting the Brownian particle.

\section{Conclusions}

We present an approach for obtaining the equilibrium distribution of a
simple system (in this case a Brownian particle under time correlated
colored noise (with Gaussian white noise as a special case) via
time-averaging (at the limit $t\rightarrow\infty$) since taking the
time-average is the most correct way of describing the long time
equilibrium behavior of any physical system.

Our model separates clearly the the irreversible contributions from
the terms that keep track of the initial conditions (memory terms). It
allows us to study each term contributing to the probability
distribution for the position and velocity of the Brownian particle
and to follow it as $t\rightarrow\infty$. It is shown, by Laplace
transforming the distributions, that the non-contributing terms
(memory terms) disappear due to the vanishing of the main propagator
associated with them in the limit $z\rightarrow 0\Leftrightarrow
t\rightarrow\infty$.  The non-vanishing terms can then be grouped
together and integrated, yielding the exact equilibrium probability
distribution, as has been confirmed by comparing with previous exact
results in the literature~\cite{caceresexato,sancho,wio}.

In summary, the main advantage of the method is that it allows us to
obtain the exact derivation of the equilibrium (and also the transient
behavior if necessary, but not done in this paper) in a systematic
way. It ''opens'' the problem into treatable parts (however ugly they
look) that reach easily calculable values at infinitely long times.

Work is under way to extend the present results to other forms of
noise.

\section{Acknowledgments}
The authors are grateful to C. Anteneodo and E.M.F. Curado for
reading and commenting the manuscript. WAMM would like to thank
Faperj and CNPq (Brazil for partial financial support and
grants. DOSP would like to thank CNPq for financial support.

\newpage

\appendix
\setcounter{equation}{0}
\section{}
The detailed Laplace transformation of Equation~\ref{Peqexp} is done below:
\begin{eqnarray}
P_{eq}(x,v)&=&\lim_{z\rightarrow0}z
\int_{0}^{\infty}e^{-zt}\langle{\delta}(x-x(t)){\delta}(v-v(t))\rangle
dt \nonumber\\
&=&\lim_{z\rightarrow0}z
\int_{0}^{\infty}dte^{-zt}\int_{-\infty}^{+\infty}\frac{dQ}{2\pi}e^{iQx}\int_{-\infty}^{+\infty}\frac{dP}{2\pi}e^{iPv}
\sum_{l=0}^{\infty}\frac{(-iQ)^{l}}{l!}\sum_{m=0}^{\infty}\frac{(-iP)^{m}}{m!}\langle
x^{l}(t)v^{m}(t)\rangle \nonumber\\
&=&\lim_{z\rightarrow0}z
\int_{0}^{\infty}dte^{-zt}\int_{-\infty}^{+\infty}\frac{dQ}{2\pi}e^{iQx}\int_{-\infty}^{+\infty}\frac{dP}{2\pi}e^{iPv}
\sum_{l=0}^{\infty}\frac{(-iQ)^{l}}{l!}\sum_{m=0}^{\infty}\frac{(-iP)^{m}}{m!} \nonumber\\
&\times &
\int_{0}^{\infty}\prod_{f=1}^{l}dt_{lf}
\int_{0}^{\infty}\prod_{h=1}^{m}dt_{mh}\delta(t-t_{la})\delta(t-t_{mb})
\langle
\prod_{f=1}^{l}
x(t_{lf})
\prod_{h=1}^{m}v(t_{mh})\rangle \nonumber\\
&=&\lim_{z\rightarrow0}z
\int_{0}^{\infty}dte^{-zt}\int_{-\infty}^{+\infty}\frac{dQ}{2\pi}e^{iQx}\int_{-\infty}^{+\infty}\frac{dP}{2\pi}e^{iPv}
\sum_{l=0}^{\infty}\frac{(-iQ)^{l}}{l!}\sum_{m=0}^{\infty}\frac{(-iP)^{m}}{m!} \nonumber\\
&\times &\int_{-\infty}^{+\infty}\prod_{f=1}^{l}\frac{dq_{f}}{2\pi}\prod_{h=1}^{m}\frac{dp_{h}}{2\pi}
\int_{0}^{\infty}\prod_{f=1}^{l}dt_{lf}
\int_{0}^{\infty}\prod_{h=1}^{m}dt_{mh} \nonumber\\
&\times&
e^{\sum_{a=1}^{l}(t-t_{la})(iq_{a}+\epsilon)
+\sum_{b=1}^{m}(t-t_{mb})(ip_{b}+\epsilon)}
\langle
\prod_{f=1}^{l}
x(t_{lf})
\prod_{h=1}^{m}v(t_{mh})\rangle \nonumber\\
&=&\lim_{z,\epsilon\rightarrow0}\sum_{l,m=0}^{\infty}
\int_{-\infty}^{+\infty}\frac{dQ}{2\pi}\frac{dP}{2\pi}e^{iQx+iPv}
\frac{(-iQ)^{l}}{l!}\frac{(-iP)^{m}}{m!}\int_{-\infty}^{+\infty}
\prod_{f=1}^{l}\frac{dq_{f}}{2\pi}\prod_{h=1}^{m}\frac{dp_{h}}{2\pi}
\nonumber\\
&\times&\frac{z}{z-\left[\sum_{f=1}^{l}iq_{f}+
\sum_{h=1}^{m}ip_{h}+(l+m)\epsilon\right]}
\langle\prod_{f=1}^{l}\tilde{x}(iq_{f}+\epsilon)\prod_{h=1}^{m}
\tilde{v}(ip_{h}+\epsilon)\rangle
\end{eqnarray}

\section{}
\begin{figure}
\epsfysize=15.0cm
\epsfxsize=12.0cm
\centerline{\rotatebox{-90}{\epsfbox{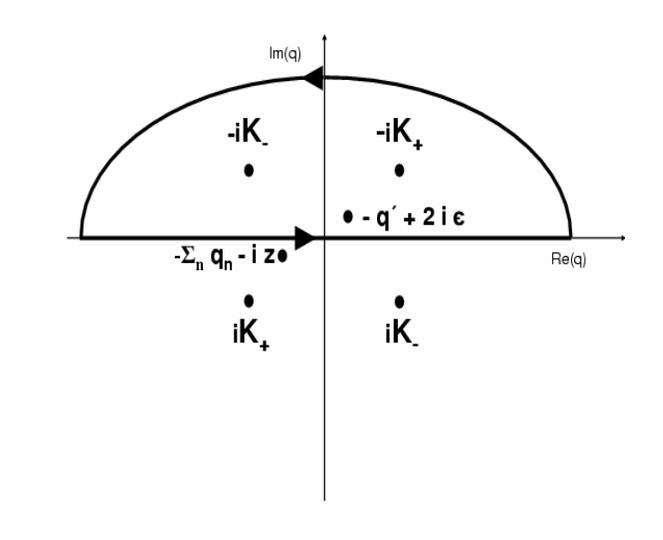}}}
\caption{Integration path for the q or p--variables.}
\label{fig1}
\end{figure}

The typical contribution for the equilibrium distribution for $x$ is
(integrations follow the path described in Figure~\ref{fig1}):

\begin{eqnarray}
&&\int_{-\infty}^{+\infty}\frac{dq_{i}}{2\pi}\frac{dq_{j}}{2\pi}
\frac{z}{z-i(q_{i}+q_{j}+\diamond)}B(iq_{i}+\epsilon)B(iq_{j}+\epsilon)
\langle\tilde{\eta}(iq_{i}+\epsilon)\tilde{\eta}(iq_{j}+\epsilon)\rangle=
\nonumber\\
&=&\int_{-\infty}^{+\infty}\frac{dq_{i}}{2\pi}\frac{dq_{j}}{2\pi}
\frac{z}{z-i(q_{i}+q_{j}+\diamond)}\frac{m^{-2}}{(q_{i}+i\kappa_{+})
(q_{i}+i\kappa_{-})(q_{j}+i\kappa_{+})(q_{j}+i\kappa_{-})}
\frac{D}{[i(q_{i}+q_{j})+2\epsilon]}\nonumber\\
&=&
\frac{z}{z-i(0+\diamond)}
\int_{-\infty}^{+\infty}\frac{dq_{j}}{2\pi}
\frac{m^{-2}D}{[q_{j}-i\kappa_{+}][q_{j}-i\kappa_{-}]
(q_{j}+i\kappa_{+})(q_{j}+i\kappa_{-})}\nonumber\\
&=&\frac{D}{2\gamma k} \frac{z}{z-i(0+\diamond)}
\end{eqnarray}

\newpage

\end{document}